\documentclass{aip-cp}
\pdfoutput=1
\usepackage[numbers]{natbib}
\usepackage{rotating}
\usepackage{graphicx}
\usepackage[T1]{fontenc}
\usepackage{currvita}

\begin{document}

\title{Shedding New Light on Kaon-Nucleon/Nuclei Interaction and Its Astrophysical Implications with the AMADEUS Experiment at DA$\Phi$NE}

\author[aff1]{A.~Scordo\corref{cor1}}
\author[aff1]{M.~Bazzi}
\author[aff2]{G.~Bellotti}
\author[aff3]{C.~Berucci}
\author[aff4]{D.~Bosnar}
\author[aff5]{A.M.~Bragadireanu}
\author[aff1]{A.~Clozza}
\author[aff3]{M.~Cargnelli}
\author[aff1]{C.~Curceanu}
\author[aff2]{A.~Dawood Butt}
\author[aff1]{R.~Del Grande}
\author[aff6]{L.~Fabbietti}
\author[aff2]{C.~Fiorini}
\author[aff1]{F.~Ghio}
\author[aff1]{C.~Guaraldo}
\author[aff1]{M.~Iliescu}
\author[aff1]{P.~Levi Sandri}
\author[aff3]{J.~Marton}
\author[aff5]{D.~Pietreanu}
\author[aff1,aff7]{K.~Piscicchia}
\author[aff1]{H.~Shi}
\author[aff1,aff5]{D.~Sirghi}
\author[aff1,aff5]{F.~Sirghi}
\author[aff8]{I.~Tucakovic}
\author[aff4]{O.~Vazquez Doce}
\author[aff3]{W.~Wiedmann}
\author[aff3]{J.~Zmeskal}

\affil[aff1]{INFN Laboratori Nazionali di Frascati, Frascati (Roma), Italy.}
\affil[aff2]{Politecnico di Milano, Dipartimento di Elettronica, Informazione e Bioingegneria and INFN Sezione di Milano, Milano, Italy.}
\affil[aff3]{Stefan-Meyer-Institut f\"ur subatomare Physik, Vienna, Austria.}
\affil[aff4]{Physics Department, University of Zagreb, Zagreb, Croatia.}
\affil[aff5]{Horia Hulubei National Institute of Physics and Nuclear Engineering (IFIN-HH), Magurele, Romania.}
\affil[aff6]{Excellence Cluster Universe, Technische Universit\"at M\"unchen, Garching, Germany.}
\affil[aff7]{Museo Storico della Fisica e Centro Studi e Ricerche "Enrico Fermi", Roma, Italy.}
\affil[aff8]{Ru\dj er Bo\u{s}kovi\'c Institute, Bijeni\u{c}ka cesta 54, Zagreb, Croatia.}
\corresp[cor1]{Corresponding author: alessandro.scordo@lnf.infn.it}

\maketitle

\begin{abstract}
The AMADEUS experiment deals with the investigation of the low-energy kaon-nuclei hadronic interaction at the DA$\Phi$NE collider at LNF-INFN, which is fundamental to respond longstanding questions in the non-perturbative QCD strangeness sector. The antikaon-nucleon potential is investigated searching for signals from possible bound kaonic clusters, which would open the possibility for the formation of cold dense baryonic matter. The confirmation of this scenario may imply a fundamental role of strangeness in astrophysics. AMADEUS step 0 consisted in the reanalysis of 2004/2005 KLOE dataset, exploiting $K^-$ absorptions in H, $^4He$, $^9Be$ and $^{12}C$ in the setup materials. In this paper, together with a review on the multi-nucleon $K^-$ absorption and the particle identification procedure, the first results on the $\Sigma^0$p channel will be presented including a statistical analysis on the possible accomodation of a deeply bound state.
\end{abstract}

\section{INTRODUCTION}

The AMADEUS experiment \cite{AMADE,AMADEUS} deals with the study of the low-energy interactions of the negatively charged kaons with light nuclei.
Such type of physics, extremely important for the understanding of the non-perturbative  QCD in the strangeness sector, has important consequences, going from hadron and nuclear physics to astrophysics. In this context, useful information can be obtained from the strength of the $K^-$ binding in nuclei. 
The investigation of the absorptions of $K^-$ inside the KLOE Drift Chamber (DC)
was originally motivated by the prediction of the formation of deeply bound
kaonic nuclear states \cite{wycech,AkYam}.  Their binding energies and widths
could be determined by studying their decays into hyperons and nucleons. Also intimately connected with the kaon-nucleon potential is the $\Lambda(1405)$ resonance, of which the still puzzling nature can be investigated within AMADEUS \cite{Hyodo}. 
The study of the KN interaction at low energies is of interest not only for quantifying the meson-baryon potential with strange content, but also because of its impact on models describing the structure of neutron stars (NS) \cite{NS}. The KN potential is attractive, as theory predicts \cite{KNtheo} and kaonic atoms confirm \cite{SID}, and this fact leads to the formulation of hypotheses about antikaon role inside the dense interior of neutron stars; one or more nucleons could be kept together by the strong attractive interaction between antikaons and nucleons and the so-called kaonic bound states, as $ppK^-$ or $ppnK^-$, might be formed. The observation of such states and the measurements of their binding energies and widths, would provide a quantitative measurement of the KN interaction in vacuum, representing an important reference for the investigation of the in-medium properties of kaons.
From the experimental point of view, two main
approaches have been used for studying the
$K^- pp$ cluster:
$p$-$p$ and heavy ion collisions
\cite{fopi} \cite{hades1},
and in-flight or stopped $K^-$ interactions
in light nuclei.
For the second, results have been published by the FINUDA \cite{finuda}
and KEK-PS E549 collaborations \cite{kek}.
The interpretation of both results is far from being conclusive,
and it requires an accurate description
of the single and multi-nucleon absorption processes that a
$K^-$ would undergo when interacting with light nuclei.
From the analysis of the KLOE 2004-2005, information on both the strength of the $K^-$ binding in nuclei and the in-medium modification of the $\Sigma^*$ and $\Lambda^*$ resonances properties can be extracted by analysing, respectively, the $\Lambda/\Sigma-p,d,t$ channels and at the resonances decay channels $\Lambda/\Sigma-\pi$.
In this paper we focus, in particular, on the analysis of the $\Sigma^0p$ final state produced in absorption processes of $K^-$ on two or more nucleons, occurring in the KLOE DC entrance wall, and on the search for a signature of the $ppK^- \rightarrow \Sigma^0+p$ kaonic bound state. A more detailed description of the analysis procedures can be found in \cite{ARXIV}.

\section{The DA$\Phi$NE collider and the KLOE detector}

DA$\Phi$NE \cite{dafne} (Double Anular $\Phi$-factory for Nice Experiments) is a double ring $e^+ \, e^-$ collider, designed to work at the center of mass energy of the $\phi$ particle $m_\phi = (1019.456 \pm 0.020) \, MeV/c^2$.
The $\phi$ meson decay produces charged kaons (with BR($K^+ \, K^-$) = $48.9 \pm 0.5 \%$) with low momentum ($\sim 127$ $MeV/c$) which is ideal either to stop them, or to explore the products of the low-energy nuclear absorptions of $K^-$s.  
The KLOE detector  \cite{kloe} is centered around the interaction region of DA$\Phi$NE and is characterised by a $\sim 4\pi$ geometry and an acceptance of $\sim98\%$; it consists of a large cylindrical Drift Chamber (DC) and a fine sampling
lead-scintillating fibers calorimeter, all immersed in an axial magnetic field of 0.52 T, provided by a superconducting solenoid.
The DC \cite{kloedc} has an inner radius of 0.25 m, an outer radius of 2 m and a length of 3.3 m. The DC entrance wall composition is 750 $\mu m$ of carbon fibre and 150 $\mu m$ of aluminium foil.
Dedicated GEANT MonteCarlo simulations  of the KLOE apparatus were performed to estimate the percentages of $K^-$ absorptions in the materials of the DC entrance wall (the $K^-$ absorption physics were treated by the GEISHA package). Out of the total number of kaons interacting in the DC entrance wall, about 81$\%$ results to be absorbed in the carbon fibre component and the residual 19$\%$ in the aluminium foil.
The KLOE DC is filled with a mixture of helium and isobutane (90$\%$ in volume $^4$He and 10$\%$ in volume $C_4H_{10}$). 
The chamber is characterised by excellent position and momentum re\-solutions. 
Tracks are reconstructed with a resolution in the transverse $R-\phi$ plane of
$\sigma_{R\phi}\sim200\,\mu m$ and a resolution along the z-axis of $\sigma_z\sim2\,mm$.
The transverse momentum resolution for low momentum tracks ($(50<p<300) MeV/c$)
is $\frac{\sigma_{p_T}}{p_T}\sim0.4\%$.
The KLOE calorimeter \cite{kloeemc} is composed of a cylindrical barrel and two endcaps,
providing a solid angle coverage of 98\%.
The volume ratio (lead/fibres/glue=42:48:10) is optimised for
a high light yield and a high efficiency for photons in the range
(20-300) $MeV/c$. The position of the cluster along the fibres can be obtained with a resolution $\sigma_{\parallel} \sim 1.4\, cm/\sqrt{E(GeV)}$. The resolution in the orthogonal direction is  $\sigma_{\perp} \sim 1.3\, cm$. The energy and time resolutions for photon clusters are given by $\frac{\sigma_E}{E_\gamma}= \frac{0.057}{\sqrt{E_\gamma (GeV)}}$ and 
$\sigma_t= \frac{57 \, ps}{\sqrt{E_\gamma (GeV)}} \oplus 100 \,\, ps$.
As a step 0 of AMADEUS, we analysed the 2004-2005 KLOE collected data, for which the $dE/dx$ information of the reconstructed tracks is available ($dE/dx$ represents the truncated mean of the ADC collected counts due to the ionisation in the DC gas). An important contribution of in-flight $K^-$ nuclear captures, in different nuclear targets from the KLOE materials, was evidenced and characterised, enabling to perform invariant mass spectroscopy of in-flight $K^-$ nuclear captures \cite{bormio}.

\section{Preliminary results of the data analyses}\label{prelim}

The investigation of the negatively charged kaons interactions in nuclear matter is performed through the reconstruction of hyperon-pion and hyperon-nucleon/nucleus correlated pairs productions, following the $K^-$ absorptions in H, ${}^4$He, ${}^9$Be and ${}^{12}$C.
The investigation of the $K^-$ multi-nucleon absorptions and the properties of possible antikaon multi-nucleon bound states proceeds through the analyses of the $\Lambda/\Sigma-p,d,t$, correlations; this last channel is, in particular, extremely promising for the search and characterisation in different nuclear targets of the extremely rare four nucleon absorption process. 
The search for the $\Lambda(1405)$ is performed through its decay into $\Sigma^0\pi^0$ (purely isospin I=0) and $\Sigma^+\pi^-$ (also the analysis of the $\Sigma^-\pi^+$ decay channel started recently with  a characterisation of neutron clusters in the KLOE calorimeter). The line shapes of the three combinations $(\Sigma\pi)^0$ were recently obtained, for the first time, in a photoproduction experiment \cite{moriya}; as the line-shapes of the three invariant mass spectra were found to be different, a comparative study with $K^-N$ production is of extreme interest. Moreover, a precise measurement of the $\frac{\Sigma^+\pi^-}{\Sigma^-\pi^+}$ production ratio in different targets can unveil the nature of the $\Lambda^*$ state, by observing modifications of its parameters as a function of the density  \cite{wycech1,ohnishi}.  
To conclude, given the excellent resolution for the  $\Lambda \pi^-$ invariant mass, the analysis of the $\Lambda \pi^-$ (isospin I=1) production, both from direct formation process and from internal conversion of a primary produced $\Sigma$ hyperon ($\Sigma \, N \rightarrow \Lambda \, N'$) is presently ongoing. Our aim is to measure, for the first time, the module of the non-resonant transition amplitude (compared with the resonant $\Sigma^{*-}$) below threshold.

\subsection{The $\Lambda(1116)$ selection}\label{lambda}

\begin{figure}[ht]
\centering

\begin{tabular}{rl}
\hspace{-1.cm}
\mbox{\includegraphics*[width=7cm]{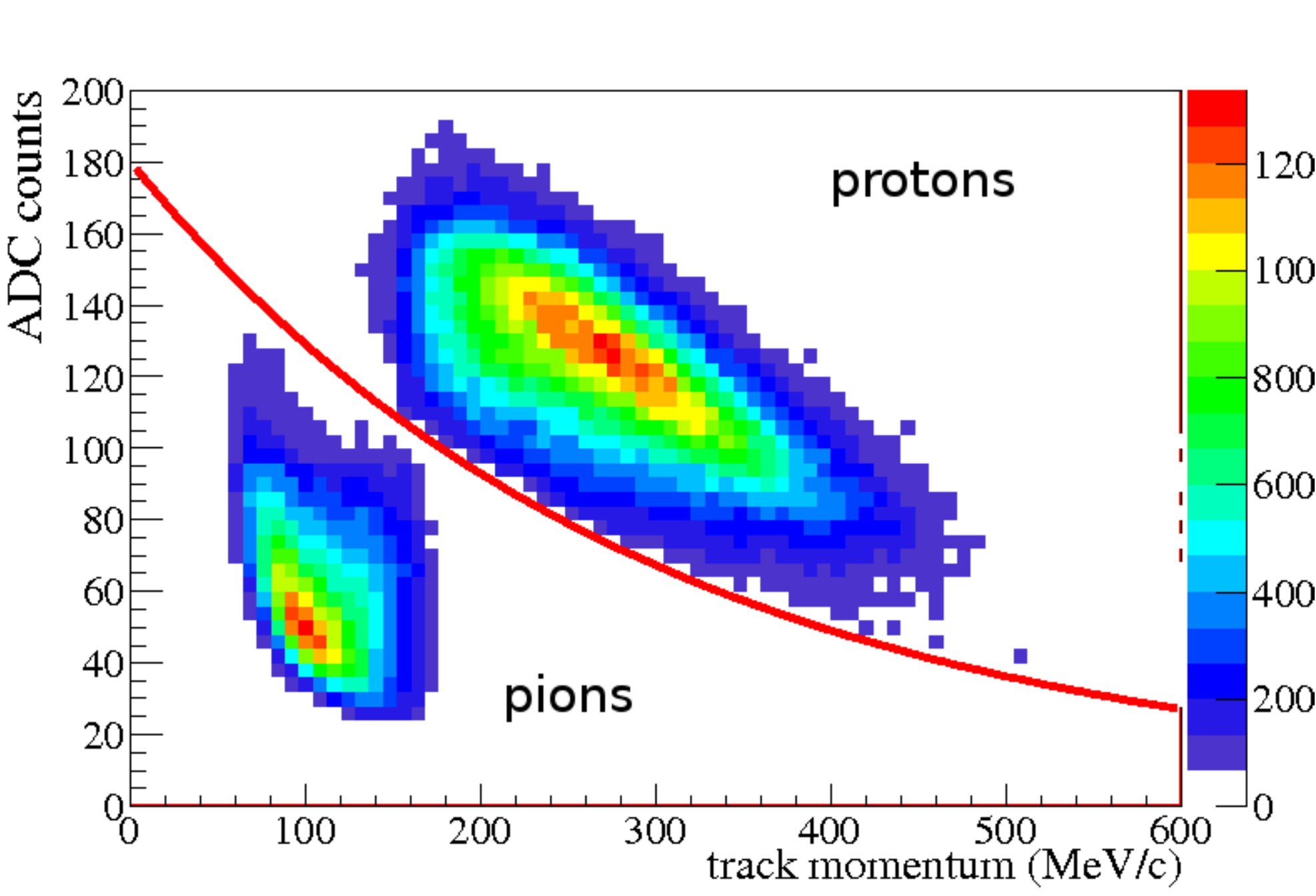}}
\hspace{-.1cm}
\mbox{\includegraphics*[width=6cm]{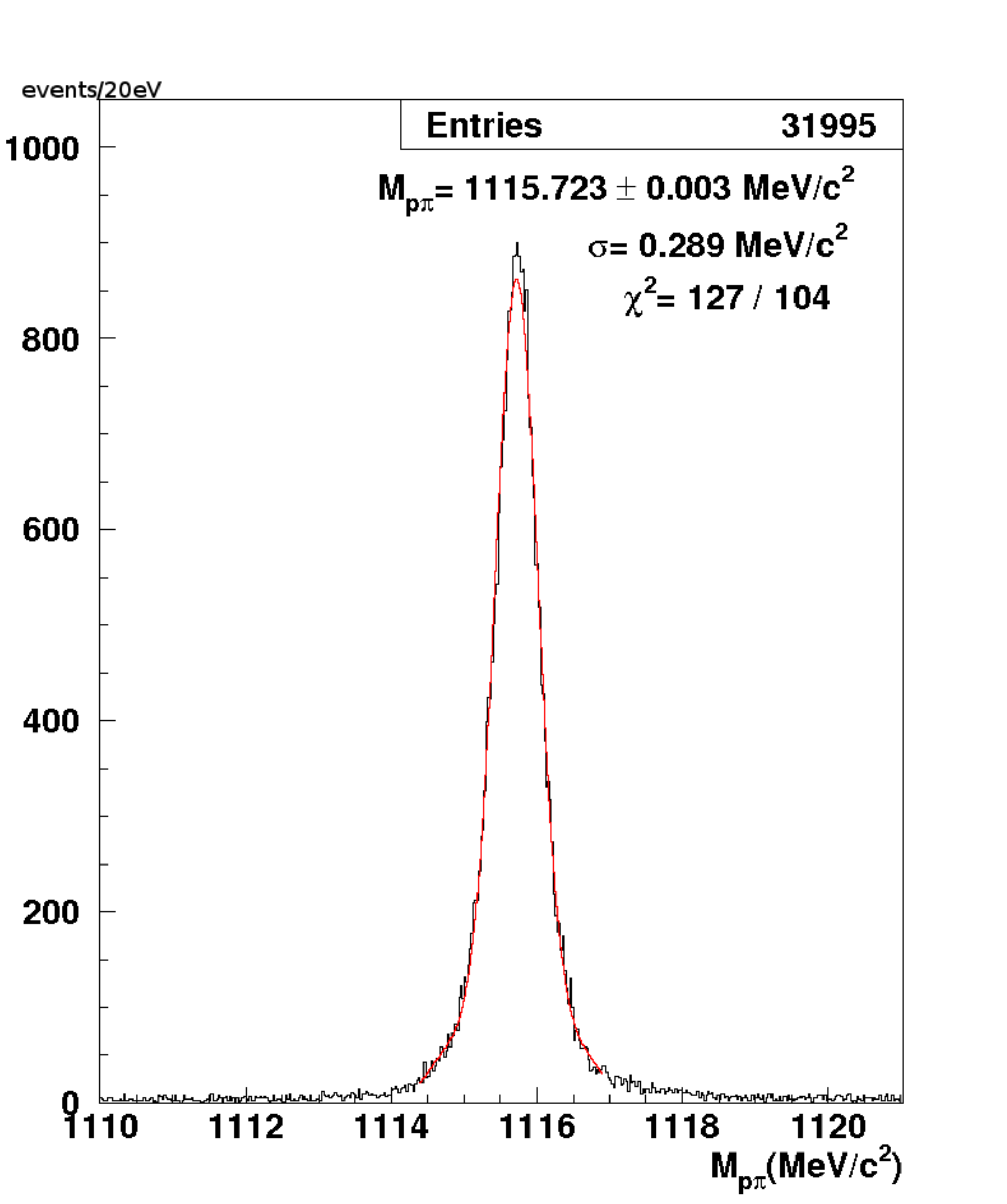}}
\end{tabular}


\caption{\em Left: $dE/dx$ (in ADC counts) vs. momentum for the selected proton (up) and pion (down) tracks in the final selection. The proton selection functions is displayed in red. Right: $m_{p\pi^-}$ invariant mass spectrum for the selected pion-proton pairs.}

\label{Lmass}       
\end{figure}

The presence of a hyperon always represents the signature of a $K^-$ hadronic interaction inside the KLOE setup materials. Most of the analyses introduced in the previous section then start with the identification of a $\Lambda(1116)$, through the reconstruction of the $\Lambda \rightarrow p + \pi^-$ (BR = 63.9 $\pm 0.5 \%$) decay vertex.
In figure \ref{Lmass} left
the $dE/dx$ versus momentum scatterplot for the finally selected protons
is shown, where the function used for the selection of protons is displayed in red.
 The typical signature of pions in $dE/dx$ versus momentum can be also seen in figure \ref{Lmass} left illustrating the efficient rejection of 
$\pi^+$ contamination in a broad range of momentum.
A minimum track length of 30 cm is required, and a common vertex is searched for all the $p-\pi^-$ pairs in each event.
When found, the common vertex position is added as an additional constraint for the track refitting. The module of the momentum and the vector cosines are redefined for both tracks,
taking into account for the energy loss in the gas and the various crossed materials (signal and field wires, DC wall, beam pipe) when tracks are extrapolated back through the detector.
As a final step for the identification of $\Lambda$ decays, the vertices are cross checked with quality cuts using the minimum distance between tracks (minimum distance $<$ 3.2 cm) and the chi-square of the vertex fit.
A spatial resolution below 1 mm is achieved for vertices found inside the DC volume (evaluated with Monte Carlo).
The invariant mass $m_{p\pi^-}$, calculated under the $p$ and  $\pi^-$ mass
hypothesis, is shown in figure \ref{Lmass} right. The Gaussian fit gives a mass of 1115.723 $\pm$ 0.003 $MeV/c^2$ and an excellent resolution ($\sigma$) of 0.3 $MeV/c^2$, confirming the unique performances of KLOE for charged particles (the systematics, depending on the momentum calibration of the KLOE setup, are presently under evaluation). 

\noindent This is the common procedure for all the analysed channels starting with the initial search of a $\Lambda$ as a signature for the hadronic interaction of the $K^-$.
Cuts on the radial position ($\rho_\Lambda$) of the $\Lambda$ decay vertex were optimised in order to separate the two samples of $K^-$ absorption events occurring in the DC wall and the DC gas: $\rho_\Lambda = 25 \pm 1.2 $ cm and $\rho_\Lambda > 33$ cm, respectively. The $\rho_\Lambda$ limits were set based on MC simulations and a study of the $\Lambda$ decay path. In particular, the $\rho_\Lambda = 25 \pm 1.2 $ cm cut guarantees, for the first sample, a contamination of $K^-$ interactions in gas as low as $(5.5^{+1.3}_{-1.8} \%)$.

\section{$\Sigma^0p$ analysis}

\begin{figure}[htbp]
\centering
\includegraphics*[width=11cm]{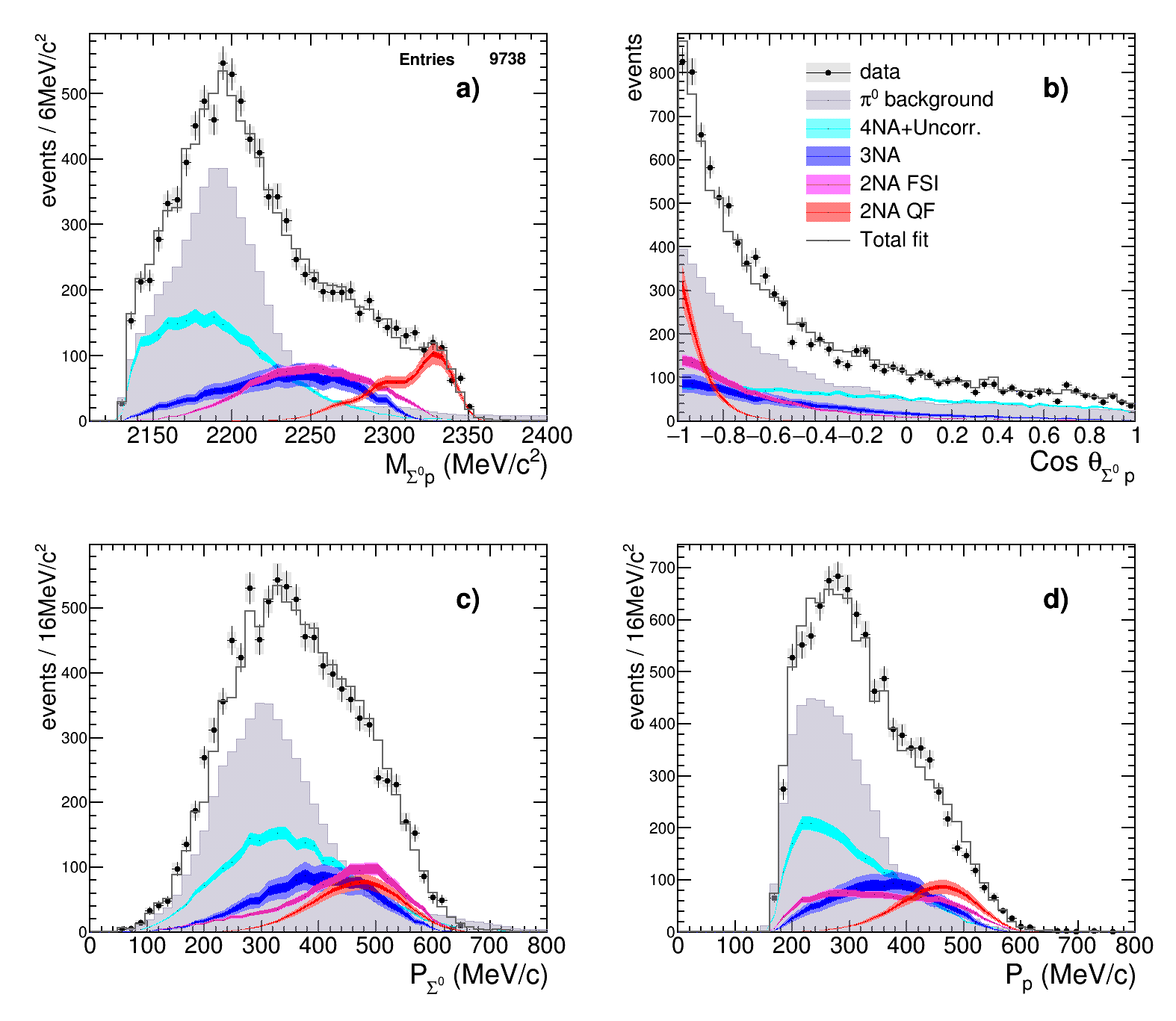}
\caption{Experimental distributions of the $\Sigma^0p$ invariant mass, $cos(\theta_{\Sigma^0p}), \Sigma^0$ and proton momentum together with the results of the global fit. The experimental data after the subtraction of the machine background are shown by the black circles, the systematic errors are represented by the boxes and the coloured histograms correspond to the fitted signal distributions where the light-coloured bands show the fit errors and the darker bands represent the symmetrised systematic errors. The gray line show the total fit distributions (see \cite{ARXIV} for details).} 
\label{fitnobs}       
\end{figure}

After the $\Lambda$ search, a common vertex between the $\Lambda$ candidate and an additional proton track is searched for. The obtained resolution on the radial coordinate for the $\Lambda p$ vertex is 12 mm, while its invariant mass resolution is found to be, from MC studies, equal to $1.1 \, MeV/c^2$.
The $\Sigma^0$ candidates are identified through their decay into $\Lambda\gamma$ pairs. After the reconstruction of a $\Lambda p$ pair, the photon selection is carried out via its identidication in the EMC. Photon candidates are selected by applying a cut on the difference between the EMC time measurement and the expected time of arrival of the photon within $-1.2<\Delta t<1.8\,ns$. Then, the $\Sigma^0p$ invariant mass, opening angle, and the individual $\Sigma^0$ and proton momenta distributions are considered simultaneously in a global fit to extract the contributions of the various absorption processes.
The processes that are taken into account in the fit of the experimental data are:

\begin{itemize}
\item $K^-A \rightarrow \Sigma^0-(\pi)p_{spec}(A')$
\item $K^-pp \rightarrow \Sigma^0-p (2NA)$
\item $K^-ppn \rightarrow \Sigma^0-p-n (3NA)$
\item $K^-ppnn \rightarrow \Sigma^0-p-n-n (4NA)$
\end{itemize}

\noindent where A is the atomic number of the target nucleus, $p_{spec}$ is the spectator proton, A' is the atomic number of the residual nucleus and 2/3/4NA stands for 2/3/4-nucleons absorption.
This list includes the $K^-$ absorption on two nucleons with and without final state interaction (FSI) for the $\Sigma^0p$ state and processes involving more than two nucleons in the initial state.
These contributions are either extracted from experimental data samples or modelled via simulations. 
Two kinds of background contribute to the analysed $\Sigma^0p$ final state: the machine background and the events with $\Lambda\pi^0p$ in the final state. Both are quantified using experimental data \cite{ARXIV}.
The obtained fit is shown in figure \ref{fitnobs} and the results are summarised in table \ref{tabfitnobs}.

\begin{table}[htbp]
\centering
\begin{tabular}{cccc}
\hline
Process & yield / $K^-_{stop}\times10^{-2}$ & $\sigma_{stat}\times10^{-2}$ & $\sigma_{syst}\times10^{-2}$ \\
\hline
2NA-QF     & $0.127$ & $\pm0.019$ & $^{+0.004}_{-0.008}$ \\
2NA-FSI    & $0.272$ & $\pm0.028$ & $^{+0.022}_{-0.023}$ \\
Tot 2NA    & $0.376$ & $\pm0.033$ & $^{+0.023}_{-0.032}$ \\
3NA        & $0.274$ & $\pm0.069$ & $^{+0.044}_{-0.021}$ \\
Tot 3 body & $0.546$ & $\pm0.074$ & $^{+0.048}_{-0.033}$ \\
4NA + bkg. & $0.773$ & $\pm0.053$ & $^{+0.025}_{-0.076}$ \\
\end{tabular}
\label{tabfitnobs}       
\caption{Production probability of the $\Sigma^0p$ final state for
different intermediate processes normalised to the number of
stopped $K^-$ in the DC wall. The statistical and systematic
errors are shown as well \cite{ARXIV}.}
\end{table}

\begin{figure}[htbp]
\centering
\includegraphics*[width=11cm]{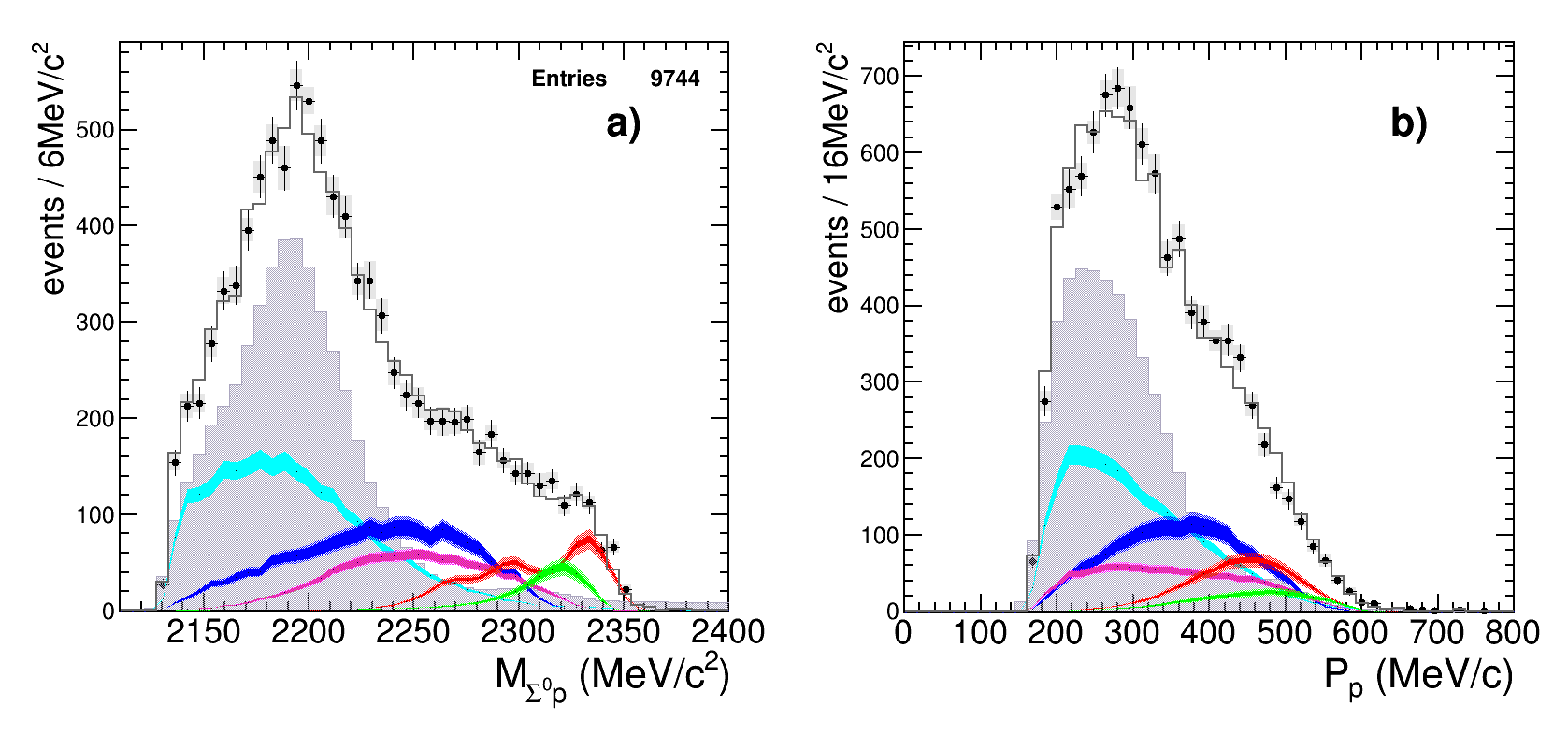}
\caption{$\Sigma^0p$ invariant mass and proton momentum distributions together with the results of the global fit
including the $ppK^-$. The different contributions are labeled as in figure \ref{fitnobs} and the green histograms represent the $ppK^-$ signal}
\label{fitbs}       
\end{figure}

\noindent The final fit results deliver the contributions of the different channels to the analysed $\Sigma^0p$ final state. The best fit
delivers a $\chi^2$ of 0.85. The emission rates extracted from
the fit are normalised to the total number of stopped antikaons.
The fit results lead
to the first measurements of the genuine 2NA-QF for the
final state $\Sigma^0p$ in reactions of stopped $K^-$ on targets of
$^{12}C$ and $^{27}Al$. This contribution is found to be only 12\%
of the total absorption cross-section.
The last step of the analysis consists in the search
of the $ppK^-$ bound state produced in $K^-$ interaction
with nuclear targets, decaying into a $\Sigma^0p$ pair. The
$ppK^-$ are simulated similarly to the 2NA-QF process
but sampling the mass of the $ppK^-$ state with a Breit-Wigner distribution, rather than the Fermi momenta of
the two nucleons in the initial state. The event kinematic is obtained by imposing the momentum conservation of the $ppK^-$ residual nucleus system. Different
values for the binding energy and width varying within
$15-75 \,MeV/c^2$ and $30-70 \,MeV/c^2$ in steps of 15
and 20 $MeV/c^2$ , respectively, are tested. This range
has been selected according to several theoretical predictions
present in literature and taking into account the experimental resolution. The global fit is repeated adding the $ppK^-$. The best fit ($\chi^2/ndf= 0.807$) is obtained for a $ppK^-$ candidate with a binding energy of
$45\, MeV/c^2$ and a width of $30\, MeV/c^2$, respectively. 
Figure \ref{fitbs} shows the results of the best fit for the $\Sigma^0p$
 invariant mass and proton momentum distributions where
the $ppK^-$ bound state contribution is shown in green.

\noindent The resulting yield normalised to the number of stopped $K^-$ is $ppK^-/K^-_{stop}=(0.044\pm0.009stat^{+0.004}_{-0.005}syst)\times10^{-2}$.

\noindent The F-test conducted to compare the simulation models with and without the 
$ppK^-$ signal gave a significance of the
result of only 1$\sigma$ for the
$ppK^-$ yield result \cite{ARXIV}. This shows that although the measured spectra are compatible with the hypothesis of a
contribution of a deeply bound state, the significance of the result is not sufficient to claim the discovery of this state.

\section{Conclusions and perspectives}

The broad experimental program of AMADEUS, dealing with the non-perturbative QCD in the strangeness sector, is supported by the quest for high precision and statistics measurements, able to set more stringent constraints on the existing theoretical models. We demonstrated the capabilities of the KLOE detector to perform high quality physics (taking advantage of the unique features of the DA$\Phi$NE factory) in the open sector of strangeness nuclear physics. Our investigations, presently spread on a wide spectrum of physical processes, represent the most ambitious and systematic effort in this field.
In particular, in this report we have presented the analysis of the $K^-$ absorption
processes leading to the $\Sigma^0$p state measured with the
KLOE detector. It was shown that the full kinematics of
this final state can be reconstructed and a global fit of
the kinematic variables allows to pin down quantitatively
the various contributing processes. 
Also, the possibility to accomodate a signal from a $ppK^-$ bound state has been 
investigated.
We proved the possibility to deliver very accurate and valuable results in the
stangeness sector, in particular for what concerns the comprehension of the KN potential, 
by studying all the possible channels following a $K^-$ absorption on one or several nucleons, for very low momentum kaons. 
For the future, a dedicated AMADEUS setup, with dedicated gaseous and solid targets, where to enhance the fraction of stopped kaons, is under study.

\section{ACKNOWLEDGMENTS}

We thank all the KLOE Collaboration and the DA$\Phi$NE staff for the fruitful collaboration. 

\noindent We acknowledge the Croatian Science Foundation under Project No. 1680.

\noindent Part of this work was supported
by the European Community-Research Infrastructure Integrating Activity ``Study of Strongly 
Interacting Matter'' (HadronPhysics2, Grant Agreement No. 227431, and HadronPhysics3 (HP3)
Contract No. 283286) under the EU Seventh Framework Programme.


\end{document}